\documentclass[a4paper,12pt]{article}
\usepackage{amsmath,amsfonts,amsthm}
\usepackage{graphics}

\begin{document}
\title{Product-limit estimators of the gap time distribution of a
renewal process under different sampling patterns}

\author{Richard D. Gill\\
Department of Mathematics\\
University of Leiden\\
\\
Niels Keiding\\
Department of Biostatistics\\
University of Copenhagen
}

\date{1 February 2010}
\maketitle
\begin{abstract}
\noindent 
Nonparametric estimation of the gap time distribution in a simple renewal process may be 
considered a problem in survival analysis under particular sampling frames corresponding to how 
the renewal process is observed. This note describes several such situations where simple product 
limit estimators, though inefficient, may still be useful.
\end{abstract}

\section{Introduction}

This note is about two classical problems in nonparametric statistical analysis of recurrent event 
data, both formalised within the framework of a simple, stationary renewal process.

We first consider observation around a fixed time point, i.e., we observe a backward recurrence time  $R$  and a forward recurrence time  $S$.  It is well known that the nonparametric maximum likelihood 
estimator of the gap-time distribution is the Cox-Vardi estimator (Cox 1969, Vardi 1985) derived from the length-biased 
distribution of the gap time  $R+S$. However, Winter \& F\"oldes (1988) proposed to use a 
product-limit estimator based on $S$,  with delayed entry given by  $R$.  Keiding \& Gill (1988) clarified the 
relation of that estimator to the standard left truncation problem. Unfortunately this discussion was 
omitted from the published version (Keiding \& Gill, 1990). Since these simple relationships do not 
seem to be on record elsewhere, we offer them here.

The second observation scheme considers a stationary renewal process observed in a finite interval 
where the left endpoint does not necessarily correspond to an event. The full likelihood function is 
complicated, and we briefly survey possibilities for restricting attention to various partial 
likelihoods, in the nonparametric case again allowing the use of simple product-limit estimators.

\section{Observation of a stationary renewal\\ process around a fixed point}

Winter \& F\"oldes (1988) studied the following estimation problem. Consider  $n$  independent 
renewal processes in equilibrium with underlying distribution function $F$,  which we shall assume 
absolutely continuous with density $f$, minimal support interval $(0,\infty)$, and hazard 
$\beta(t)=f(t)/(1-F(t))$, $t>0$. The reason for our unconventional choice $\beta$ for the hazard rate belonging to $F$ will become apparent later. Corresponding to a fixed time, say $0$, the backward and forward recurrence 
times  $R_i$ and $S_i$, $i=1,...,n$,  are observed;  their sums $Q_i=R_i+S_i$  are length-biased observations from 
$F$, i.e., their density is proportional to $tf(t)$.  Let $(R,S,Q)$ denote a generic triple $(R_i,S_i,Q_i)$.  We quote the following distribution results: let  $\mu$  be the expectation value corresponding to the
the distribution $F$,
$$\mu~=~\int_0^\infty u f(u) \mathrm du~=~\int_0^\infty(1-F(u))\mathrm du,$$
then the joint distribution of $R$ and $S$ has density $f(r+s)/\mu$ , the marginal distributions of $R$ and $S$ 
are equal with \emph{density}  $(1-F(r))/\mu$, and the marginal distribution of $Q=R+S$  has density 
$qf(q)/\mu$, the length-biased density corresponding to  $f$.

Winter and F\"oldes considered the product-limit estimator
$$
1-\widetilde F(t)~=~\prod_{i:Q_i\le t}\biggl(1-\frac 1 {Y(Q_i)}\biggr)
$$
where
$$
Y(t)~=~\sum_{i=1}^nI\{R_i<t\le R_i+S_i\}
$$
is the \emph{number at risk} at time $t$.  This estimator is the same as the Kaplan-Meier estimator for iid 
survival data $Q_1,\dots,Q_n$ left-truncated at $R_1,\dots,R_n$  (Kaplan \& Meier 1958, Andersen et al.~1993). 
Winter \& F\"oldes showed that $1-\widetilde F$ is strongly consistent for the \emph{underlying} survival function 
$1-F$ .

We shall show how the derivation of this estimator follows from a simple Markov process model 
similar to the one used by Keiding \& Gill (1990) to study the random truncation model.

First notice that the conditional distribution of $Q=R+S$  given that $R=r$  has density
$$
\frac{ f(q)/\mu  }  {  (1-F(r))/\mu  },\quad r < q < \infty
$$	 
that is, intensity (hazard)  $f(q)/(1-F(q))$, $q>r$, which is just the hazard $\beta(q)$  corresponding to the 
underlying distribution  $F$ left-truncated at $r$.  Now define corresponding to $(R,S,Q)$ a stochastic 
process  $U$  on $[0,\infty]$  with state space $\{0,1,2\}$  by
$$
U(t)~=~\Biggl\{
\begin{aligned}
& 0, &  &~~t<~R,\\
& 1, & R~\le&~~t<~R+S,\\
&  2, & \quad R+S~\le &~~t .
\end{aligned}
$$
Note that it takes in successsion the values $0$, $1$ and $2$. For $U(t)=0$,
\begin{align*}
P\bigl\{U&(t+h)=1\bigm|U(u),0\le u\le t\bigr\}\\
~&=~P\bigl\{R \le t+h\bigm|R>t\bigr\}\\
~&=~\alpha(h)h +o(h),
\end{align*}
where $\alpha$ is the hazard rate of the marginal distribution of $R$.
For $U(t)=1$ (and hence $R\le t$)
\begin{align*}
P\bigl\{U&(t+h)=2\bigm|U(u),0\le u\le t\bigr\}\\
~&=~P\bigl\{R+S\le t+h\bigm|R=r\le t,R+S>t\bigr\}\\
~&=~\frac {f(t)} { 1-F(t) } h + o(h)
\end{align*}
by the above result on the conditional hazard of  $R+S$ given $R$. For $U(t)=0$,
$$
P\bigl\{U(t+h)=2\bigm|U(u),0\le u\le t\bigr\}~=~o(h).
$$
Other transitions are impossible.

That these conditional probabilities depend on $U(t)$ and $t$ only, but not on $U(u)$, $u<t$, proves that  $U$  
is a Markov process
with intensities
$$
\alpha(t)~=~ \frac { 1-F(t) } { \int_t^\infty (1-F(r) ) \mathrm d r }
$$
(the marginal hazard of $R$,  equal to the residual mean lifetime function of the underlying 
distribution  $F$) and
$$
\beta(t)~=~\frac {f(t) } { 1-F(t) },
$$
see Figure 1.
\begin{figure}
\centerline{\includegraphics{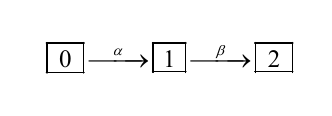}}
\caption{Inhomogenous 3-state Markov process, 2 allowed transitions}
\end{figure}
The Markov process framework of Keiding \& Gill (1990) now indicates that 
(ignoring information about $F$ in $\alpha$, and just focussing on the transition with rate $\beta$)
the product limit 
estimator $1-\widetilde F$  is a natural estimator of the survivor function $1-F$  of interest, 
and consistency and asymptotic normality may be obtained as shown by Keiding \& Gill (1990, Sec.~5).

Note that the backwards intensity
\begin{align*}
{\overline\alpha}(t)~&=~\alpha(t)\frac{  P\bigl\{ U(t)=0 \bigr\} } { P\bigl\{ U(t)=1 \bigr\} }\\
~&=~\alpha(t)\frac{ P\bigl\{ R>t \bigr\} } { P\bigl\{ R\le t<R+S \bigr\}  }\\
~&=~\alpha(t)\frac{ \mu^{-1} \int_t^\infty(1-F(r))\mathrm d r } { \mu^{-1} \int_0^t\int_{t-r}^\infty f(r+s)\mathrm d s \mathrm d r } 
\\
~&=~\frac {1-F(t) }  { \int_t^\infty (1-F(r)) \mathrm d r } \frac{ \int_t^\infty(1-F(r))\mathrm d r } { \int_0^t(1-F(t)) \mathrm d r } ~=~\frac1t ,
\end{align*}
the \emph{backwards} hazard-rate of a uniform distribution on a bounded interval  $(0,A)$, $A<\infty$. Since it has been assumed that $R$ 
has support interval $(0,\infty)$, this shows that the present model \emph{may not} 
be interpreted strictly as a left truncation 
model, which would require that ${\overline\alpha}(t)$ was the backwards hazard rate of some probability distribution on $(0,\infty)$. However, this distinction is not important to our discussion.

The fact that ${\overline\alpha}(t)$ does not depend on $F$ corresponds to Winter and F\"oldes' statement that $(R,S)$  contains no 
more information than $R+S$  about $F$. This already follows from sufficiency since the joint density of $(R,S)$ is 
$f(r+s)/\mu$. The likelihood 
function based on observation of $(R_1,S_1),\dots(R_n,S_n)$ is
$$
\mu^{-n}\prod_{i=1}^n f(r_i+s_i)
$$
from which the NPMLE of  $F$  is readily derived as
$$
\widehat F(t)~=~\sum_{i=1}^n \frac { I\bigl\{ R_i+S_i\le t\bigr\} } { R_i+S_i} \biggm / \sum_{i=1}^n\frac 1 { R_i+S_i},
$$
that is the Cox-Vardi estimator in the terminology of Winter and F\"oldes (Cox 1969, Vardi 1985).

It follows that the estimator $1-\widetilde F$  is \emph{not} NPMLE. The important difference between the 
situation here and 
that of the random truncation model studied by Keiding \& Gill (1990, Sec. 3) is that not only the 
intensity $\beta(t)$, but also $\alpha(t)$ depends only on the estimand  $F$.

As already mentioned, weak convergence of $1-\widetilde F$  is immediate from Keiding \& Gill (1990, Sec. ~5). 
In particular, in order to achieve the extension to convergence on $[0,M]$  it should be required 
that
$$
\int_0^\varepsilon	  \mathrm d \Phi(s) /\nu_2(s)~<~\infty
$$
in the terminology of Keiding \& Gill (1990, Sec.~5c), and using  $\mathrm d \Phi(t)=\beta(t)\mathrm d t$  and
$$
\nu_2(t)~=~P\bigl\{ U(t)=1\bigr\} ~=~\int_0^t\frac {1-F(s) } {\mu} \frac {1-F(t)}{1-F(s)}\mathrm d s~=\frac t\mu (1-F(t)),
$$ 
the integrability condition translates into
$$
\int_0^\varepsilon \frac { \beta(t)}  {P\bigl\{ U(t)=1\bigr\} } \mathrm d t~ < \infty
$$
or finiteness of  $E(1/X)$ where  $X$ has the underlying (``length-unbiased'') interarrival time 
distribution $F$. It may easily be seen from Gill et al.~(1988) that the same condition is needed to 
ensure weak convergence of the Cox-Vardi estimator.

A variation of the observation scheme of this section would be to allow also right censoring of the $S_i$. This can be immediately included in the Markov-process/counting process approach leading to the inefficient product-limit type estimator $1-\widetilde F$; the delayed-entry observations $S_i$ are simultaneously right-censored. See Vardi (1985, 1989) and Asgharian et al.~(2002) for treatment of the full non-parametric maximum likelihood estimator of $F$, extending the Cox-Vardi estimator to allow right censoring.

Other ad hoc estimators and the rich relationships with a number of other important non-parametric estimation problems are discussed by Denby and Vardi~(1985) and Vardi (1989).

\section{Observation of a stationary renewal\\ process in a finite interval}

We consider again a stationary renewal process on the whole line and assume that we observe it in 
some interval $[t_1,t_2]$ determined independently of the process. Nonparametric estimation of the gap 
time distribution  $F$  was definitively discussed by Vardi (1982) in discrete time and by Soon \& 
Woodroofe~(1996) in continuous time. Cook \& Lawless (2007, Chapter~4) surveyed the general area 
of analysis of gap times emphasizing that the assumption of independent gap times is often 
unrealistic.

We shall here nevertheless work under the assumption of the simplest possible model as indicated 
above. Because the nonparametric maximum likelihood estimator is computationally involved it 
may sometimes be useful to calculate less efficient alternatives, and there are indeed such 
possibilities.

Under the observation scheme indicated above we may have the following four types of elementary 
observations
\\

\noindent 1.	Times $x_i$ from one renewal to the next, contributing the density $f(x_i)$  to the likelihood.
\\

\noindent  2.	Times from one renewal  $T$  to  $t_2$, which are right-censored versions of 1., 
contributing factors of the form $(1-F(t_2-T))$  to the likelihood.
\\

\noindent 3.	Times from $t_1$  to the first renewal $T$ (forward recurrence times), contributing factors of the 
form $(1-F(T-t_1))/\mu$   to the likelihood.
\\

\noindent 4.	Knowledge that no renewal happened in $[t_1,t_2]$ , actually a right-censored version of 3., 
contributing factors of the form $\int_{t_2-t_1}^\infty (1-F(u))\mathrm d u/\mu$  to the likelihood.
\\

McClean \& Devine (1995) studied nonparametric maximum likelihood estimation in the 
conditional distribution given that there is at least one renewal in the interval, i.e., that there are no 
observations of type 4.

Our interest is in basing the estimation only on complete or right-censored gap times, i.e., 
observations of type 1 or 2. When this is possible, we have simple product-limit estimators in the 
one-sample situation, and we may use well-established regression models (such as Cox regression) 
to account for covariates. Pe\~na et al.~(2001) assumed that observation started at a renewal (thereby 
defining away observations of type 3 and 4) and gave a comprehensive discussion of exact and 
asymptotic properties of product-limit estimators with comparisons to alternatives, building in 
particular on results of Gill (1980, 1981) and Sellke (1988). The crucial point here is that calendar 
time and time since last renewal both need to be taken into account, so the straightforward 
martingale approach displayed by Andersen et al.~(1993) is not available. Pe\~na et al.\ also studied 
robustness to deviations from the assumption of independent gap times.

As noted by Aalen \& Husebye (1991) in their attractive non-technical discussion of observation 
patterns, observation does however often start between renewals. (In the example of Keiding et al.\  
(1998), auto insurance claims were considered in a fixed calendar period). As long as observation 
starts at a stopping time, inference is still valid, so by starting observation at the first renewal in the 
interval we can essentially refer back to Pe\~na et al.~(2001). A more formal argument could be based 
on the concept of the \emph{Aalen filter}, see Andersen et al.~(1993, p.~164). The resulting product-limit 
estimators will not be fully efficient, since the information in the backward recurrence time (types 3 
and 4) is ignored. It is important to realize that the validity of this way of reducing the data depends 
critically on the independence assumptions of the model. Keiding et al.~(1998), cf.\ Keiding (2002) 
for details, used this fact to base a goodness-of-fit test on a comparison of the full nonparametric 
maximum likelihood estimator with the product-limit estimator.

Similar terms appear in another model, called the \emph{Laslett line segment problem} (Laslett, 1982). 
Suppose one has a stationary Poisson process, with intensity $\mu$, of points on the real line. We 
think of the real line as a calendar time axis, and the points of the Poisson process will be called 
\emph{pseudo renewal times} or \emph{birth times} of some population of individuals. Suppose the individuals 
have independent and identically distributed lifetimes, each one starting at the corresponding 
birth time. The corresponding calendar time of the end of each lifetime can of course be called a 
\emph{death time}. Now suppose that \emph{all we can observe} are the intersections of individuals' lifetimes 
(thought of as time segments on the time axis) with an observational window $[t_1,t_2]$.  In 
particular, we do not know the current age of an individual who is observed alive at time  $t_1$.  
Again we have exactly the same four kinds of observations: 
\\

\noindent 1.	Complete \emph{proper} lifetimes corresponding to births within  $[t_1,t_2]$  for which death 
occurred before time $t_2$.
\\

\noindent 2.	Censored \emph{proper} lifetimes corresponding to births within $[t_1,t_2]$ for which death 
occurred after time  $t_2$.
\\

\noindent 3.	Complete \emph{residual} lifetimes corresponding to births which occurred at an 
unknown moment before time $t_1$, and for which death occurred after $t_1$  and before time $t_2$.
\\
 
\noindent 4.	Censored \emph{residual} lifetimes corresponding to births which occurred at an 
unknown moment before time $t_1$, for which death occurred after time $t_2$, and which are 
therefore censored at time  $t_2$.
\\

\noindent The \emph{number} of at least partially observed lifetimes (proper or residual) is random, and Poisson 
distributed with mean equal to the intensity $\mu$  of the underlying Poisson process of birth times, 
times the factor  
$$
t_2-t_1+\int_0^\infty (1 - F(y)) \mathrm d y .
$$
This provides a fifth, ``Poisson'', factor in the nonparametric likelihood function for parameters 
$\mu$  and $F$, based on all the available data. Maximizing over $\mu$ and $F$, the mean of the Poisson 
distribution is estimated by the observed number of partially observed lifetimes. Thus we find 
that the \emph{profile likelihood} for $F$, and the \emph{marginal likelihood} for $F$ based only on contributions 
1.--4., are proportional to one another.

Nonparametric maximum likelihood estimation of $F$ was studied by Wij\-ers~(1995) and van der 
Laan (1996), cf.\ van der Laan \& Gill (1999). Some of their results, and the calculations leading 
to this likelihood, were surveyed by Gill (1994, pp.~190~ff.). The nonparametric maximum 
likelihood estimator is consistent; whether or not it converges in distribution as $\mu$  tends to 
infinity is unknown, the model has a singularity coming from the vanishing probability density of 
complete lifetimes just larger than the length of the observation window corresponding to births 
just before the start of the observation window and deaths just after its end. Van der Laan 
showed that a mild reduction of the data by grouping or binning leads to a much better behaved 
nonparametric maximum likelihood estimator. If the amount of binning decreases at an 
appropriate rate as  $\mu$ increases, this leads to an asymptotically efficient estimator of $F$. This 
procedure can be thought of as \emph{regularization}, a procedure often needed in nonparametric 
inverse statistical problems, where maximum likelihood can be too greedy.

Both ``unregularized'' and regularized estimators are easy to compute with the EM algorithm; 
and the speed of the algorithm is not so painfully slow as in other inverse problems, since this is 
still a problem where ``root $n$'' rate estimation is possible.

The problem allows, just as we have seen in earlier sections, all the same inefficient but rapidly 
computable product-limit type estimators based on various marginal likelihoods. Moreover 
since the direction of time is basically irrelevant to the model, one can also look at the process 
``backwards'', leading to another plethora of inefficient but easy estimators. One can even 
combine in a formal way the censored survival data from a forward and a backward time point 
of view, which comes down to counting all uncensored observations twice, all singly censored 
once, and discarding all doubly censored data. (This idea was essentially suggested much earlier 
by R.C.~Palmer and D.R.~Cox, cf.\ Palmer(1948)). The attractive feature of this estimator is 
again the ease of computation, the fact that it only discards the doubly censored data, and its 
symmetry under reversing time. The asymptotic distribution theory of this estimator is of course 
not standard, but using the nonparametric delta method one can fairly easily give formulas for 
asymptotic variances and covariances. In practice one could easily and correctly use the 
nonparametric bootstrap, resampling from the partially observed lifetimes, where again a 
resampled complete lifetime is entered twice into the estimate. 

The Laslett line segment problem has rather important extensions to observation of line 
segments (e.g., cracks in a rock surface) observed through an observational window in the 
plane. Under the assumption of a homogenous Poisson line segment process one can write 
down nonparametric likelihoods, maximize them with the EM algorithm; it seems that 
regularization may well be necessary to get optimal ``root $n$'' behaviour but in principle it is 
clear how this might be done. Again, we have the same plethora of inefficient but easy 
product-limit type estimators. Van Zwet (2004) studied the behaviour of such estimators when the line 
segment process is not Poisson, but merely stationary. The idea is to use the Poisson process 
likelihood as a quasi likelihood, i.e., as a basis for generating estimating equations, which will 
be unbiased but not efficient, just as in parametric quasi-likelihood. Van Zwet shows that this 
procedure works fine. Coming full circle, one can apply these ideas to the renewal process we 
first described in this section, and the other models described in earlier sections. All of them 
generate stationary line segment processes observed through a finite time window on the line. 
Thus the nonparametric quasi-likelihood approach can be used there too. Since in the renewal 
process case we are ignoring the fact that the intensity of the point process of births equals the 
inverse mean life-time, we do not get full efficiency. So it is disputable whether it is worth 
using an inefficient ad-hoc estimator which is difficult to compute when we have the options of 
Soon and Woodroofe's fully efficient (but hard to compute) full nonparametric maximum 
likelihood estimator, and the many inefficient but easy and robust product-limit type estimators 
of this paper.

\section*{Acknowledgements}
This research was partially supported by a grant (RO1CA54706-12) from the National Cancer 
Institute and by the Danish Natural Sciences Council grant 272-06-0442 ``Point process modelling 
and statistical inference''.

\section*{References}
\raggedright
\frenchspacing

Aalen, O.O. \& Husebye, E. (1991). Statistical analysis of repeated events forming renewal 
processes. \emph{Statistics in Medicine} \textbf{10}, 1227--1240.

\par~ \par
Andersen, P.K., Borgan, \O., Gill, R.D. \& Keiding, N. (1993). \emph{Statistical Models Based on Counting 
Processes}. Springer Verlag, New York.

\par~ \par
Asgharian, M., Wolfson, D. B. \&  M'lan, C. E. (2002). Length-biased sampling with right censoring: an unconditional approach. \emph{J.\ Amer.\ Statist.\ Assoc.}  \textbf{97}, 201--209.

\par~ \par
Cook, R.J. \& Lawless, J.F. (2007). \emph{The statistical analysis of recurrent events}. Springer Verlag, 
New York.

\par~ \par
Cox, D.R. (1969). Some sampling problems in technology. In \emph{New De\-vel\-opments in Survey 
Sampling} (N.L.~Johnson and H.~Smith, Jr., eds), 506--527. Wiley, New York.

\par~ \par
Denby, L. \& Vardi, Y. (1985). A short-cut method for estimation in renewal processes. \emph{Technometrics} \textbf{27}, 361--373.

\par~ \par
Gill, R.D. (1980). Nonparametric estimation based on censored observations of a Markov renewal 
process. \emph{Zeitschrift f\"ur Wahr\-schein\-lich\-keits\-theorie und verwandte Gebiete} \textbf{53}, 97-116.

\par~ \par
Gill, R.D. (1981). Testing with replacement and the product-limit estimator. \emph{Annals of Statistics} \textbf{9}, 
853--860.

\par~ \par
Gill, R.D. (1994). Lectures on survival analysis. \emph{Lecture notes in Mathematics} \textbf{1581}, 115--241. 
Springer, New York.

\par~ \par
Gill, R.D., Vardi, Y. \& Wellner, J.A. (1988). Large sample theory of empirical distributions in 
biased sampling models. \emph{Annals of Statistics} \textbf{16}, 1069--1112.

\par~ \par
Kaplan, E.L. \& Meier, P. (1958). Non-parametric estimation from incomplete observations. \emph{Journal 
of the American Statistical Association}~\textbf{53}, 457--481.

\par~ \par
Keiding, N. (2002). Two nonstandard examples of the classical stratification approach to 
graphically assessing proportionality of hazards. In: \emph{Goodness-of-fit Tests and Model Validity} (eds. 
C.~Huber-Carol, N.~Balakrishnan, M.S.~Nikulin and M.~Mesbah). Boston, Birkh\"auser, 301--308.

\par~ \par
Keiding, N., Andersen, C. \& Fledelius, P. (1998). The Cox regression model for claims data in non-life 
insurance. \emph{ASTIN Bulletin} \textbf{28}, 95--118.

\par~ \par
Keiding, N. \& Gill, R.D. (1988). \emph{Random truncation models and Markov processes}. Report MS-
R8817, Centre for Mathematics and Computer Science, Amsterdam.

\par~ \par
Keiding, N. \& Gill, R.D. (1990). Random truncation models and Markov processes. \emph{Annals of 
Statistics} \textbf{18}, 582--602.

\par~ \par
Laslett, G.M. (1982). The survival curve under monotone density constraints with application to 
two-dimensional line segment processes. \emph{Biometrika} \textbf{69}, 153--160.

\par~ \par
McClean, S. \& Devine, C. (1995). A nonparametric maximum likelihood estimator for incomplete 
renewal data. \emph{Biometrika} \textbf{82}, 791--803.

\par~ \par
Palmer, R.C. (1948). The dye sampling method of measuring fibre length distribution (with an 
appendix by R.C.~Palmer and D.R.~Cox). \emph{Journal of the Textile Institute} \textbf{39}, T8--T22. 

\par~ \par
Pe\~na, E. A., Strawderman, R. L. \& Hollander, M. (2001). Nonparametric estimation with recurrent 
event data. \emph{Journal of the American Statistical Association} \textbf{96}, 1299--1315.

\par~ \par
Sellke, T. (1988). Weak convergence of the Aalen estimator for a censored renewal process, in 
\emph{Statistical Decision Theory and Related Topics IV}, (eds.\ S.~Gupta \& J.~Berger), Springer-Verlag, 
New York, Vol. 2, 183--194.

\par~ \par
Soon, G.  \& Woodroofe, M. (1996). Nonparametric estimation and consistency for renewal processes. \emph{Journal of Statistical Planning and Inference} \textbf{53}, 171--195. 

\par~ \par
van der Laan, M.J. (1996). Efficiency of the NPMLE in the line-segment problem. \emph{Scandinavian 
Journal of  Statistics} \textbf{23}, 527--550.

\par~ \par
van der Laan, M.J. and Gill, R.D. (1999). Efficiency of NPMLE in nonparametric missing data 
models. \emph{Mathematical Methods in Statistics} \textbf{8}, 251--276.

\par~ \par
van Zwet, E.W. (2004). Laslett's line segment problem. \emph{Bernoulli} \textbf{10},  377--396.

\par~ \par
Vardi, Y. (1982). Nonparametric estimation in renewal processes. \emph{Annals of Statistics} \textbf{10}, 
772--785.

\par~ \par
Vardi, Y. (1985). Empirical distributions in selection bias models. \emph{Annals of Statistics} \textbf{13}, 
178--203.

\par~ \par
Vardi, Y. (1989). Multiplicative censoring, renewal processes, deconvolution and decreasing density: nonparametric estimation. \emph{Biometrika} \textbf{76}, 751--761

\par~ \par
Wijers, B.J. (1995). Consistent nonparametric estimation for a one-dimensional line-segment 
process observed in an interval. \emph{Scandinavian Journal of  Statistics} \textbf{22}, 335--360.

\par~ \par
Winter, B.B. \& F\"oldes, A. (1988). A product-limit estimator for use with length-biased data.  
\emph{Canadian Journal of Statistics} \textbf{16}, 337--355.

\end{document}